\documentclass[12pt]{article}

\usepackage{graphicx}

\begin{document}

\input{epsf.sty}

\begin{titlepage}

\begin{flushright}
IUHET-499\\
\end{flushright}
\vskip 2.5cm

\begin{center}
{\Large \bf Vacuum Cerenkov Radiation in Lorentz-Violating Theories Without CPT
Violation}
\end{center}

\vspace{1ex}

\begin{center}
{\large Brett Altschul\footnote{{\tt baltschu@indiana.edu}}}

\vspace{5mm}
{\sl Department of Physics} \\
{\sl Indiana University} \\
{\sl Bloomington, IN 47405 USA} \\

\end{center}

\vspace{2.5ex}

\medskip

\centerline {\bf Abstract}

\bigskip

In theories with broken Lorentz symmetry, Cerenkov radiation may be possible even in
vacuum. We analyze the Cerenkov emissions that are associated with the least
constrained Lorentz-violating modifications of the photon sector, calculating
the threshold energy, the frequency spectrum, and the shape of the Mach cone. In
order to obtain sensible results for the total power emitted, we must make use of
information contained within the theory which indicates at what scale new physics
must enter.

\bigskip

\end{titlepage}

\newpage

In recent years, there has been a great deal
of interest in the possibility of Lorentz violation, since
string theory and many
other candidate theories of quantum gravity may predict deviations from Lorentz
invariance in certain regimes. If Lorentz violation were to be observed
experimentally, it would be a discovery of momentous importance and a profound
clue regarding the structure of the universe at the most fundamental level.
Experimental searches for Lorentz violation---which have thus far not yielded any
compelling positive
results---have included studies of matter-antimatter asymmetries for
trapped charged particles~\cite{ref-bluhm1,ref-bluhm2,ref-gabirelse,
ref-dehmelt1} and bound state systems~\cite{ref-bluhm3,ref-phillips},
determinations of muon properties~\cite{ref-kost8,ref-hughes}, analyses of
the behavior of spin-polarized matter~\cite{ref-kost9,ref-heckel2},
frequency standard comparisons~\cite{ref-berglund,ref-kost6,ref-bear,ref-wolf},
Michelson-Morley experiments with cryogenic resonators~\cite{ref-antonini,
ref-stanwix,ref-herrmann}, Doppler effect measurements~\cite{ref-saathoff,ref-lane1},
measurements of neutral meson
oscillations~\cite{ref-kost10,ref-kost7,ref-hsiung,ref-abe,ref-link,ref-aubert},
polarization measurements on the light from distant galaxies~\cite{ref-carroll1,
ref-carroll2,ref-kost11,ref-kost21}, analyses of the radiation emitted by energetic
astrophysical sources~\cite{ref-jacobson1,ref-altschul6}, and others.

Significant work has also been done on the theoretical side.
A Lorentz- and CPT-violating effective field theory,
the standard model extension (SME), has been developed in detail~\cite{ref-kost1,
ref-kost2,ref-kost12}. The theory's stability,
causality~\cite{ref-kost3}, and one-loop renormalizability~\cite{ref-kost4}, have
all been examined. Recent work has also probed the question of how generic
Lorentz violation is within quantum field theory~\cite{ref-altschul5}.
The SME contains coefficients parameterizing all
possible observer-independent Lorentz violations. The results of experimental
Lorentz tests can be used to place bounds on the coefficients of the minimal SME,
which contains only gauge invariant and renormalizable parameters. Some of the
minimal SME coefficients are extremely tightly bounded, but the bounds on many
other coefficients are weak or even nonexistent.

Scattering and decay processes may be affected in unexpected ways by
Lorentz violation.
One especially interesting process
is vacuum Cerenkov radiation, $e^{-}\rightarrow e^{-}\gamma$, which is forbidden in
Lorentz-symmetric theories, since the speeds of charged particles are always less
than the speed of light propagation. Since vacuum Cerenkov radiation
could be an extremely
important energy loss process for the highest energy particles~\cite{ref-coleman3,
ref-jacobson3,ref-gagnon}, a better understanding of this process is needed.
We shall look at this kind of radiation in detail, in the presence of one particular
type of photon-sector Lorentz
violation---the type for which the experimental bounds are the weakest.
For definiteness, we shall consider a matter sector with a single fermion field,
so that the Lagrange density, omitting the Lorentz violation, is
\begin{equation}
{\cal L}_{0}=-\frac{1}{4}F^{\mu\nu}F_{\mu\nu}+\bar{\psi}[\gamma^{\mu}(i\partial_{\mu}
-eA_{\mu})-m]\psi.
\end{equation}
However, the detailed structure of the matter sector is actually
unimportant; our calculations would still be valid if the charged particles were
bosons.

Lorentz-violating
modifications of the electromagnetic sector fall into two categories---those which
generate photon birefringence and those which do not. A CPT-odd Chern-Simons
term ${\cal L}_{AF}=\frac{1}{2}k_{AF}^{\mu}\epsilon_{\mu\nu\rho\sigma}F^{\nu\rho}
A^{\sigma}$
will always generate birefringence, as will ten of the nineteen independent
coefficients in the CPT-even ${\cal L}_{F}=-\frac{1}{4}k_{F}^{\mu\nu\rho\sigma}
F_{\mu\nu}F_{\rho\sigma}$. Birefringence has been ruled out very strongly by
polarization measurements made on photons that have traversed cosmological
distances~\cite{ref-carroll1,ref-carroll2,ref-kost11,ref-kost21}. The remaining
nine coefficients, which are contained in the symmetric, traceless $k_{F\alpha}
\,^{\mu\alpha\nu}$, are much less strongly constrained, at the $10^{-16}$ level
or worse
(compared to $10^{-32}$ or better for the birefringent terms). Therefore, if we are
interested in looking for potential signatures of actual Lorentz violation in the
photon sector, it is most natural to look for effects associated with these
particular coefficients.

Ordinary Cerenkov radiation occurs when a charged particle moving in
a medium exceeds the speed of light in that medium. Something similar can occur in
the vacuum if there is Lorentz violation. When a charged particle is moving faster
than the photon signal speed in a given direction, we expect the charge to radiate.
This radiation field has been studied using both microscopic~\cite{ref-kaufhold}
and macroscopic~\cite{ref-lehnert1,ref-lehnert2} electrodynamics, for the situation
in which the source of the Lorentz violation is a $k_{AF}$.
However,
Cerenkov radiation in the presence of $k_{F}$ has not been studied in the same
detail.

If the ten components of $k_{F}$ which generate birefringence are set to zero, then
$k_{F}$ takes the form
\begin{equation}
\label{eq-kF}
k_{F}^{\mu\nu\rho\sigma}=\frac{1}{2}\left(g^{\mu\rho}k_{F\alpha}\,^{\nu\alpha\sigma}
-g^{\mu\sigma}k_{F\alpha}\,^{\nu\alpha\rho}
-g^{\nu\rho}k_{F\alpha}\,^{\mu\alpha\sigma}
+g^{\nu\sigma}k_{F\alpha}\,^{\mu\alpha\rho}\right).
\end{equation}
$\tilde{k}^{\mu\nu}\equiv k_{F\alpha}\,^{\mu\alpha\nu}$ is symmetric and traceless in
$(\mu,\nu)$. It is invariant under both C and PT.
Since $\tilde{k}$ should be small, we shall
evaluate expressions only to leading order in this parameter.

We shall exploit a duality between the theory with a $k_{F}$ as in (\ref{eq-kF})
and a different Lorentz-violating theory. The original theory has Lorentz violation
in the electromagnetic sector only; the fermion sector is conventional.
A coordinate transformation
$x^{\mu}\rightarrow
x'^{\mu}=x^{\mu}-\frac{1}{2}\tilde{k}^{\mu}\,_{\nu}x^{\nu}$ moves all
the Lorentz violation into the matter sector~\cite{ref-kost17}.
To leading order, the transformation of the Lagrange density is
\begin{equation}
{\cal L}_{0}+{\cal L}_{F}\rightarrow {\cal L}_{0}-
\frac{1}{2}\tilde{k}^{\mu\nu}\bar{\psi}\gamma_{\nu}(i\partial_{\mu}
-eA_{\mu})\psi.
\end{equation}
The $\tilde{k}$ becomes a $c$ term in the fermion sector.
If the initial theory contained Lorentz violations in both sectors, the effective $c$
would simply be the sum of the initial fermionic
$c$ and the induced $c$ coming from the gauge sector.
However, we shall neglect this possibility for simplicity.
In what we shall call the ``original'' coordinates,
all the Lorentz violation is in the photon sector; and we shall
refer to the coordinates in which the Lorentz violation has been moved entirely
into the matter sector as the ``primed'' coordinates, although we shall not write
the primes explicitly.
Most of the tightest bounds
on the combined electron $c$ and photon $\tilde{k}$ coefficients come from
observations of astrophysical synchrotron and inverse Compton radiation, combined
with the lack of observed vacuum Cerenkov radiation~\cite{ref-altschul6}.

If the Lorentz violation is all in the photon sector, the physical picture is as
follows.
The speed of electromagnetic wave propagation is generally direction dependent, and
it may be smaller than one. Then a charge moving with a velocity very close to one
may be moving faster than the physical speed of light in the same direction, and
Cerenkov radiation results. In the dual theory, in the primed coordinates,
the fermions' maximum speeds may be greater than one
in certain directions.  It is easier to consider this version, with the charges
moving superluminally,
because when the electromagnetic sector is conventional, standard results for
Cerenkov emission may be applied directly.

However, ordinary Cerenkov radiation in matter is emitted only up to some cutoff
frequency.  Above that frequency, the dielectric constant becomes close enough to
one that the signal speed exceeds the charge's speed.
This effect ensures that the total power radiated by the charge is finite. Something
similar occurs when the Cerenkov radiation induced by a $k_{AF}$
Chern-Simons term is
considered; yet with a $k_{F}$ that is independent of frequency, there is no such
cutoff, and the radiated power would appear to diverge. This is a significant but
not insurmountable complication. As we shall see, the theory with $k_{F}$ quite
naturally contains some information about the scale at which new physics must come
into play, and this will largely allow us to resolve the issue.

\begin{figure}[t]
\epsfxsize=3in
\begin{center}
\leavevmode
\epsfbox[200 300 590 650]{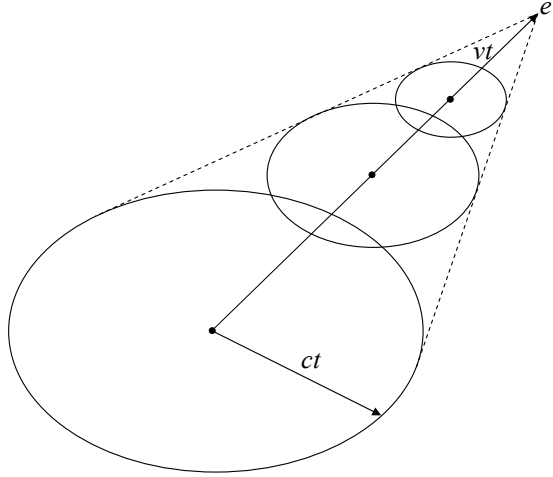}
\caption{Geometry of Cerenkov shocks in a theory with a direction-dependent speed
of light. The charge $e$ is moving with speed $v$, and the 
the ellipses represent the signal fronts. $c$ denotes the signal speed in
the indicated direction.
\label{fig-cone}}
\end{center}
\end{figure}

The basic situation---in the original coordinates---is shown in
Figure~\ref{fig-cone}. The picture is similar
to the conventional one describing Cerenkov radiation, except that the speed of light
is not isotropic. The envelope of the ellipses forms the Mach cone---the wave
front of the emitted radiation.
In general, the cone is neither right angled nor circular; the opening angles
on either side of the charge's trajectory depend on the Lorentz violation.
However,
the figure is exaggerated, in that we expect the deviation of the light speed from
one in any given direction to be a very small correction, of ${\cal O}\left(
\tilde{k}\right)$; and the smallness of the $\tilde{k}$ will simplify the situation
significantly. In the primed coordinates, where the electromagnetic sector is
conventional, the figure would be different; the signal fronts become circles, and
the shocks arise because the charge's speed is greater than one.

Since $\tilde{k}$ is small, any particle emitting Cerenkov radiation must have
a velocity $\vec{v}$ with magnitude very
close to one. (In what follows, $\vec{v}$ and $v$ will always denote
the velocity and speed in the original coordinates where the
Lorentz violation is purely electromagnetic.)
When the Lorentz violation is moved entirely into the fermion sector (so that
we are using primed coordinates), the maximum
particle speed in a direction $\hat{e}$ is $1+\frac{1}{2}\left[
\tilde{k}_{jk}\hat{e}_{j}\hat{e}_{k}+\tilde{k}_{(0j)}\hat{e}_{j}+\tilde{k}_{00}
\right]$,
where $\tilde{k}_{(0j)}=\tilde{k}_{0j}+
\tilde{k}_{j0}$. The speed $v$ becomes
$v+\frac{1}{2}\left[\tilde{k}_{jk}\hat{v}_{j}\hat{v}_{k}+\tilde{k}_{(0j)}
\hat{v}_{j}+\tilde{k}_{00}\right]$,
where $\hat{v}$ is a unit vector in the direction of $\vec{v}$, and
we have neglected the deviation of $v$ from one in the
explicitly $\tilde{k}$-dependent terms. The condition for Cerenkov emission is
therefore that $1-v<\frac{1}{2}\left[\tilde{k}_{jk}\hat{v}_{j}\hat{v}_{k}+
\tilde{k}_{(0j)}\hat{v}_{j}+\tilde{k}_{00}\right]$. At high energies, the Lorentz
factor is
$\gamma\approx1/\sqrt{2(1-v)}$, so the
energy threshold for vacuum Cerenkov radiation is
\begin{equation}
\label{eq-threshold}
E_{T}=\frac{m}{\sqrt{\tilde{k}_{jk}\hat{v}_{j}\hat{v}_{k}+
\tilde{k}_{(0j)}\hat{v}_{j}+\tilde{k}_{00}}}.
\end{equation}
If the square root in (\ref{eq-threshold}) is imaginary, the speed of light in the
direction $\hat{v}$ is greater than one, and charges moving in that direction will
never emit Cerenkov radiation.

In the primed coordinates, the Cerenkov angle $\theta_{C}$ (the angle between
$\hat{v}$ and the emitted radiation) is given by the standard formula
$\cos\theta_{C}=\left\{v+\frac{1}{2}\left[\tilde{k}_{jk}\hat{v}_{j}\hat{v}_{k}+
\tilde{k}_{(0j)}\hat{v}_{j}+\tilde{k}_{00}\right]\right\}^{-1}$, or
\begin{equation}
\theta_{C}^{2}=\tilde{k}_{jk}\hat{v}_{j}\hat{v}_{k}+
\tilde{k}_{(0j)}\hat{v}_{j}+\tilde{k}_{00}-2(1-v).
\end{equation}
Since $\theta_{C}^{2}$ is already ${\cal O}\left(\tilde{k}\right)$---$(1-v)$ being
at most ${\cal O}\left(\tilde{k}\right)$ when there is Cerenkov emission---the
Cerenkov angle is effectively the same in either set of coordinates. Thus, when
higher order corrections are neglected, the Cerenkov emission still occurs along
a right circular cone; the Mach cone is symmetric about the direction $\hat{v}$,
although the width of the cone does depend on which direction the charge is moving.
Because the speed of the moving charge can be only very
slightly greater than the signal speed, the cone is very broad. The situation is
therefore
simpler than is depicted in the exaggerated Figure~\ref{fig-cone}, and the shape of
the Mach cone is almost completely determined by the local radii of curvature of the
signal fronts where they intersect the charge's straight line path.

All the radiation is emitted at the Cerenkov angle $\theta_{C}$. The energy radiated
per unit frequency per unit time is
\begin{equation}
P(\omega)=\frac{e^{2}}{4\pi}\theta_{C}^{2}\omega
=\frac{e^{2}}{4\pi}\left[\tilde{k}_{jk}\hat{v}_{j}\hat{v}_{k}+
\tilde{k}_{(0j)}\hat{v}_{j}+\tilde{k}_{00}-2(1-v)\right]\omega.
\end{equation}
This frequency spectrum is unambiguous, at least at lower frequencies. However, at
higher frequencies, this result is somewhat problematic. Since there is a shock
front with zero thickness, the electromagnetic field contains Fourier components
at arbitrarily short wavelengths, and the total rate of energy emission appears to
diverge.  Some sort of cutoff is required if we are to obtain a sensible result.

Fortunately, the theory itself contains a natural indication of the correct cutoff
scale. The electromagnetic sector alone does not specify any energy scale; however,
the matter sector contains the mass $m$. The theory in the primed coordinates
runs into causality problems at a scale ${\cal O}\left(m\tilde{k}^{-1/2}
\right)$~\cite{ref-kost4}.
Since electromagnetic and fermionic Lorentz violations mix under
renormalization, the Lorentz-violating
electromagnetic theory coupled to fermions will fail at roughly the same scale.
Some new physics must emerge at this scale, and one of the roles they must play
will be to cut off the power spectrum $P(\omega)$ at a scale $\Lambda_{\tilde{k}}\sim
m\tilde{k}^{-1/2}$. With multiple species, the relevant mass $m$ is that of
the lightest charged particles (i.e., electrons).

Then the total power emitted becomes
\begin{equation}
\label{eq-P}
P=\frac{e^{2}m^{2}}{8\pi}\left(\frac{\theta_{C}^{2}\Lambda_{\tilde{k}}^{2}}{m^{2}}
\right),
\end{equation}
most of the emission coming around $\omega\sim\Lambda_{\tilde{k}}$.
The expression in parentheses in (\ref{eq-P}) is dimensionless and ${\cal O}(1)$.
The new physics at the cutoff scale may be Lorentz violating, so
$\Lambda_{\tilde{k}}$ could depend on $\hat{v}$. However, if Lorentz violation is
small at low energy scales, it is likely to be small at high scales
also~\cite{ref-collins}, so the dominant
contribution to $\Lambda_{\tilde{k}}$ may be
direction independent.
Not surprisingly, the threshold energy $E_{T}$ is at the
same scale as $\Lambda_{\tilde{k}}$. Since the new physics only appear at energies
above $\Lambda_{\tilde{k}}$, any particle emitting vacuum Cerenkov radiation must be
at least that energetic, so that the new physics which are necessary to make the
total emission finite can come into play.

Back reaction on the charge is relatively simple. The charge loses energy at the
rate $P$, and because $\theta_{C}$ is small, all the Cerenkov photons are beamed
into a narrow pencil of angles around $\hat{v}$. The particle therefore loses
energy and momentum at essentially the same rate, in accordance with its
ultrarelativistic dispersion relation. As the energy falls close to $E_{T}$, the
emission rate slackens, and the particle slows to the terminal speed
$v=1-\frac{1}{2}\left[\tilde{k}_{jk}\hat{v}_{j}\hat{v}_{k}+
\tilde{k}_{(0j)}\hat{v}_{j}+\tilde{k}_{00}\right]$.

At energies far above the threshold for emission, we may neglect $(1-v)$,
and the Cerenkov angle approaches $\theta_{C}^{2}=\tilde{k}_{jk}\hat{v}_{j}
\hat{v}_{k}+\tilde{k}_{(0j)}\hat{v}_{j}+\tilde{k}_{00}$. According to (\ref{eq-P}),
the rate of
energy loss becomes independent of energy, and the lifetime of a particle with
energy $E\gg E_{T}$ is $\tau\sim \frac{8\pi E}{e^{2}m^{2}}$. For an electron
with an energy of $10^{8}$ GeV, we find $\tau\sim 10^{-6}$ s. So particles with
energies above $E_{T}$ should lose their excess energies quite quickly.

It may seem paradoxical that the total power emitted is not suppressed by any
power of $\tilde{k}$. However, the power per unit frequency is so suppressed, and the
overall Cerenkov emission effect is strongly suppressed by the smallness of the
Lorentz violation, because as $\tilde{k}$ decreases in size, the threshold $E_{T}$
for there to be any emission at all is pushed to higher and higher energies.
So low-energy physics will be negligibly affected if $\tilde{k}$ is small enough.

The experimental bounds on the Lorentz-violating but CPT-even $\tilde{k}$ coefficients
are by far the weakest of any in the electromagnetic sector. This makes these
coefficients potentially among the most interesting in the SME. Analyses of vacuum
Cerenkov radiation in the presence of $\tilde{k}$ have previously been stymied by
the observation that the total emitted power would probably diverge. However, since
the theory automatically
contains information about the scale at which new physics must enter,
we have been able to circumvent that difficulty, at least enough to obtain order of
magnitude estimates; and with these results, it is now possible in principle
to search quantitatively for evidence of vacuum Cerenkov radiation in the emissions
from the highest-energy particles.

\section*{Acknowledgments}
The author is grateful to V. A. Kosteleck\'{y} and Q. Bailey for helpful discussions.
This work is supported in part by funds provided by the U. S.
Department of Energy (D.O.E.) under cooperative research agreement
DE-FG02-91ER40661.

\end{document}